# Negative thermal expansion and local lattice distortion in the $(Sc_{1-x}Ti_x)F_3$ and related solid solutions

Xiaojian Wang,[1] Jun Chen,[1,†] Fei Han,[1] Yang Ren,[2] Tao Wang,[1] Jinxia Deng,[1] and Xianran Xing[1]

[1]Department of Physical Chemistry, University of Science and Technology Beijing, Beijing 100083, China

[2]Argonne National Laboratory, X-Ray Science Division, Argonne, Illinois 60439, USA

[†]Corresponding Author, Electronic mail: Junchen@ustb.edu.cn

**Abstract:**

Negative thermal expansion (NTE) is unusual but important property for control of thermal expansion. In the present study, the chemical modification is utilized to engineer controllable thermal expansion in cubic NTE $ScF_3$. A broad window of the coefficient of thermal expansion (CTE, $\alpha_1$ = -1.51 ~ -3.4 × $10^{-6}$ $K^{-1}$, 300-800 K) has been achieved in $Sc_{1-x}M_xF_3$ ($M$ = Ti, Al and Ga). The long-range crystallographic structure of $(Sc_{1-x}Ti_x)F_3$ adheres to the cubic $Pm\bar{3}m$ symmetry according to the analysis of high-energy synchrotron X-ray powder diffraction. Pair distribution function (PDF) analysis of synchrotron X-ray total scattering was performed to investigate the local lattice distortion. It was found that the weakness of NTE has a close correlation with the local lattice distortion. Based on the coupled rotation model, it is presumably that this local distortion might dampen the transverse vibration of F atoms and thus reduce NTE. The present work provides a possible reference for the design of controllable NTE in open framework solids.

## 1. Introduction

Negative thermal expansion (NTE) is an interesting physical property of solid state chemistry, which is extensively applied in various functional materials.[1] A large amount of work have been conducted to explore controllable thermal expansion materials, such as $ZrW_2O_8$,[2] Invar alloys,[3] NZP family,[4] $PbTiO_3$-based ferroelectrics,[5-6] anti-perovskite nitrides,[7] $BiNiO_3$-based solid solutions,[8] and $ReO_3$-type fluorides.[9-16] Among them, $ScF_3$ exhibits an intriguing isotropic NTE behavior, due to its simple structure with cubic corner-shared octahedra. There have several studies to elucidate the NTE and physical properties through chemical modification in $ScF_3$ like $(Sc_{1-x}M_x)F_3$ ($M$ = Al,[13] Y,[14] Fe,[15] Ti,[16] etc.). It has been found there is a certain correlation between the controllable thermal expansion with the local lattice distortion in the $ScF_3$ based solid solutions.[11, 17]

In the present study, the chemically substituted $ScF_3$ solid solutions of $(Sc_{1-x}M_x)F_3$ ($M$ = Ti, Al, and Ga) have been prepared. A tunable thermal expansion was observed in these isotropic $ScF_3$-based systems. Interestingly, the coefficient of thermal expansion (CTE) of $(Sc_{1-x}M_x)F_3$ is found to show a close relationship with the degree of local lattice distortion. Such lattice distortion presumably could restrain the transverse motion of fluorine atoms, and thus weakens NTE. The present work could be helpful for the design of thermal expansion in NTE compounds through chemical modification.

## 2. Result and Discussion

The details of Experimental section are given in the Supporting Information (SI). As shown in Fig. S1, $ScF_3$ exhibits a cubic $ReO_3$-type framework structure which is composed of corner-sharing $ScF_6$ octahedra. Structure refinements have been performed to determine crystal structure of $(Sc_{1-x}Ti_x)F_3$ solid solutions, $(Sc_{0.9}Al_{0.1})F_3$, and $(Sc_{0.9}Ga_{0.1})F_3$ (Fig. 1, Fig. S3 and Fig. S4). It is found that the crystal structure of all compositions maintains cubic $Pm\bar{3}m$ symmetry, in which F atoms are located at the 3d site (0.5, 0, 0) and metal atoms at the 1a site (0, 0, 0). Fig. S2 shows lattice parameters as function of Ti content ($x$). One can see that the lattice parameter, $a$, decreases with increasing content of Ti ($x$). It is 4.0115(1) Å for $ScF_3$ and 3.9350(1) Å for $(Sc_{0.1}Ti_{0.9})F_3$. The decrease in lattice parameters is in a good agreement with the decrease in ionic radius for $Sc^{3+}$ (0.745 Å) and $Ti^{3+}$ (0.670 Å).[18]

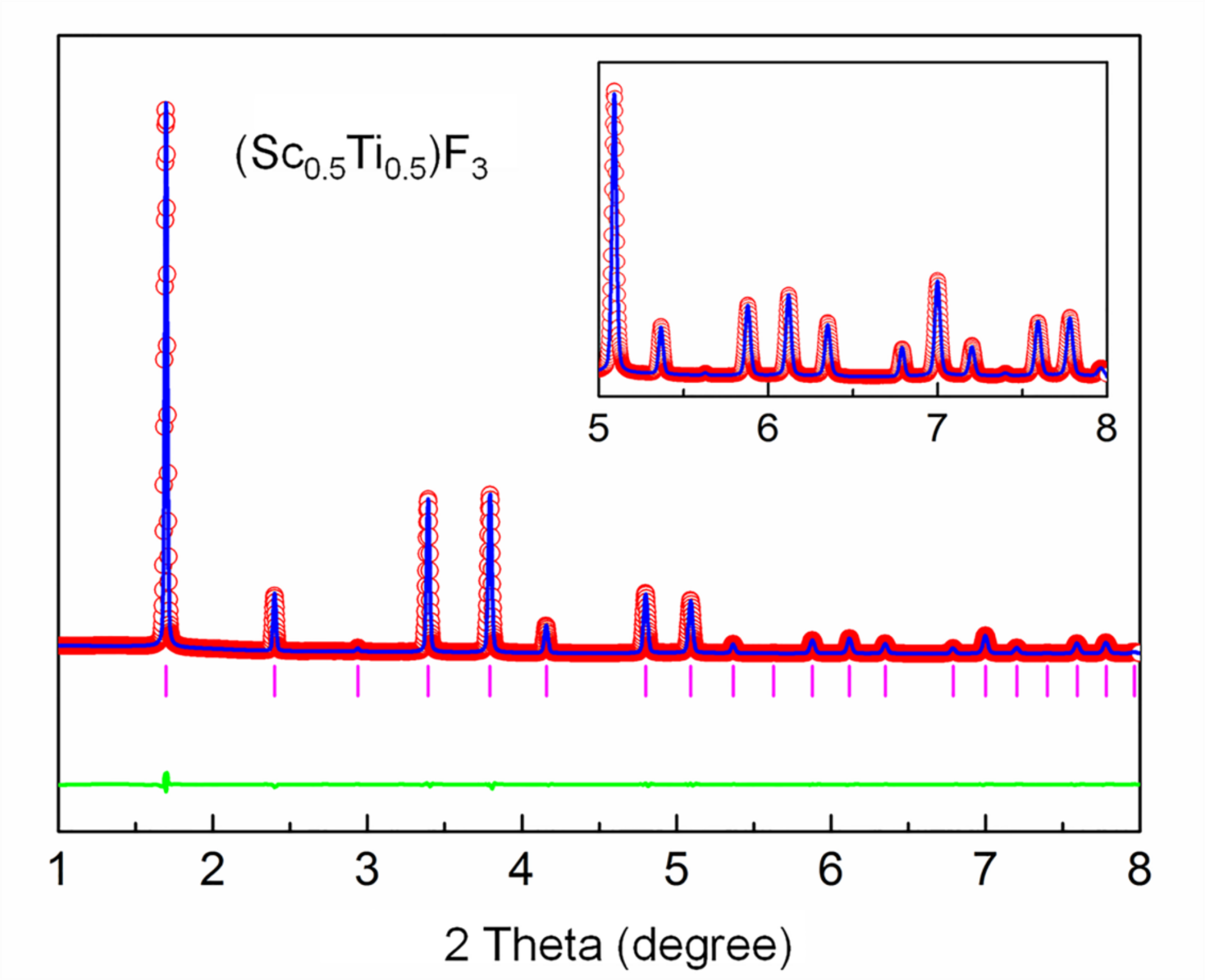

**Figure 1.** Observed, calculated, difference of patterns of structure refinement of $(Sc_{0.5}Ti_{0.5})F_3$ from high-energy synchrotron X-ray diffraction data.

As shown in Fig. 2a, the temperature dependence of lattice parameters exhibits a distinct thermal expansion as a function of $TiF_3$ content. The magnitude of NTE differs noticeably from strong NTE of $ScF_3$ to weak NTE of $(Sc_{0.1}Ti_{0.9})F_3$.

Fig. 2b provides linear CTEs of $(Sc_{1-x}Ti_x)F_3$ solid solutions as a function of Ti content ($x$). From room temperature to 800 K, $ScF_3$ exhibits a relatively strong NTE with an average linear CTE of $-3.4 \times 10^{-6}$ $K^{-1}$, which is in accordance with the value of $-3.0 \times 10^{-6}$ $K^{-1}$ in the previous study.[9] By chemical substitution of 50 mol% $Ti^{3+}$ for $Sc^{3+}$, i.e., $(Sc_{0.5}Ti_{0.5})F_3$, its NTE is weakened ($\alpha_1 = -2.94 \times 10^{-6}$ $K^{-1}$, 300-800 K).

With further chemical substitution, thermal expansion of $(Sc_{0.1}Ti_{0.9})F_3$ develops less negative ($\alpha_1$ = -1.51 $\times 10^{-6}$ $K^{-1}$, 300-800 K). As a conclusion, one can see that NTE of $(Sc_{1-x}Ti_x)F_3$ becomes more and more weakened with chemical substitution of Sc by Ti atom. Here, a nearly continuous isotropic CTEs are achieved in $(Sc_{1-x}Ti_x)F_3$. The solid solutions functionalize over a wide temperature range, especially including the high temperature up to 800K. Until now, high-temperature isotropic NTE is rare, such as $Zr_{1-x}Sn_xMo_2O_8$,[19] $TaO_2F$,[20] $MZrF_6$.[21] This present system of $(Sc_{1-x}Ti_x)F_3$ extends the diversity of high-temperature isotropic NTE species.

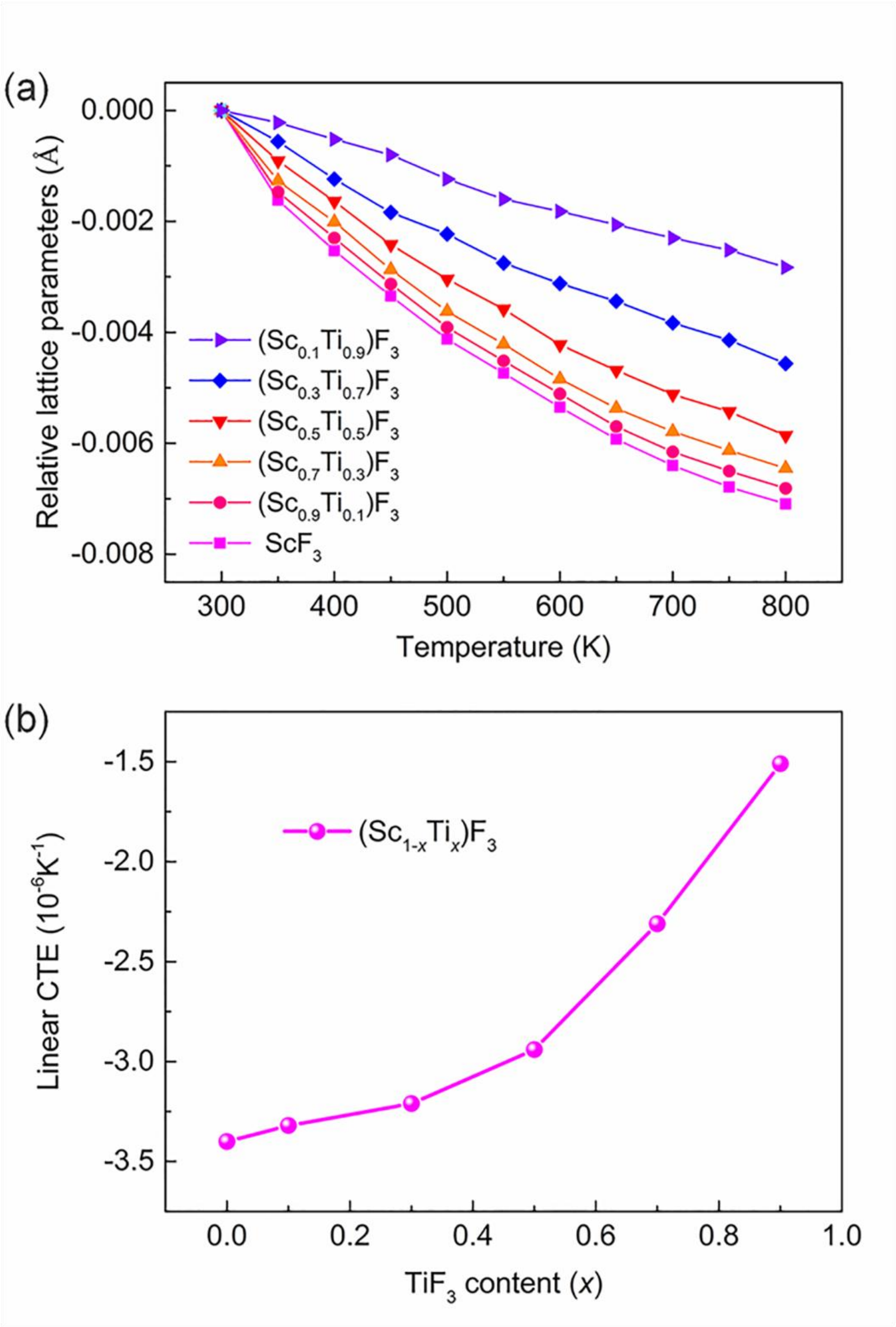

**Figure 2.** (a) Temperature-dependence variation of relative lattice parameters, and (b) linear CTE of $(Sc_{1-x}Ti_x)$ $F_3$ ($x$ = 0, 0.1, 0.3, 0.5, 0.7, and 0.9) solid solutions.

Synchrotron X-ray diffraction (SXRD) result shows that all present $ScF_3$-based solid solutions remain cubic symmetry macroscopically. If not, several apparent peaks due to symmetry breaking would appear in the SXRD patterns, like depicted in Figure S5. However, local lattice distortions could be responsible for the difference in thermal expansion. It is essential to further explore local lattice distortions. Here, we have performed the atomic pair distribution function (PDF) analysis of synchrotron radiation X-ray total scattering for the present $ScF_3$-based solid solutions (Fig. 3a, Fig. S6, and Fig. S7).

Firstly in order to be consistent with the XRD results, the cubic $Pm\bar{3}m$ model was adopted to investigate the local lattice distortion at the initial attempts. But unsatisfactorily, the fitting results indicate that the cubic model cannot match the experimental PDF data well. It could be anticipated that local lattice distortions certainly appear with the chemical substitutions of $Ti^{3+}$, $Al^{3+}$ or $Ga^{3+}$ for $Sc^{3+}$. Therefore, enlightened by the documentation that the $ReO_3$-type metal trifluorides commonly retain in $R\bar{3}c$ symmetry (such as $AlF_3$), it could be reasonable to utilize a rhombohedral $R\bar{3}c$ model to investigate the local structure. Based on this model, the refinement obtains quite acceptable fitting results. PDF analyses demonstrate that the rhombohedral model is superior to the cubic counterpart at the low r range (1.7-20 Å) (Fig. S8).

It was previously reported that $ScF_3$ adopts the cubic $ReO_3$ structure in which the Sc−F−Sc linkages are straight.[10] Related to the strong coupling rotation of $ScF_6$ octahedral units, the transverse vibration of F atom perpendicular to the straight Sc··Sc axis results in a decrease in Sc···Sc distance and thus the strong NTE.[10] However, for the distorted rhombohedral model, a certain displacement exists in F atoms deviating from its original central location and leading to bent M-F-M ( M = Ti, Al, Ga) linkages with a value of θ less than 180° (Fig. 3b). Here, the difference, $\Delta\theta = 180° - \theta$, is defined as the distortion degree.

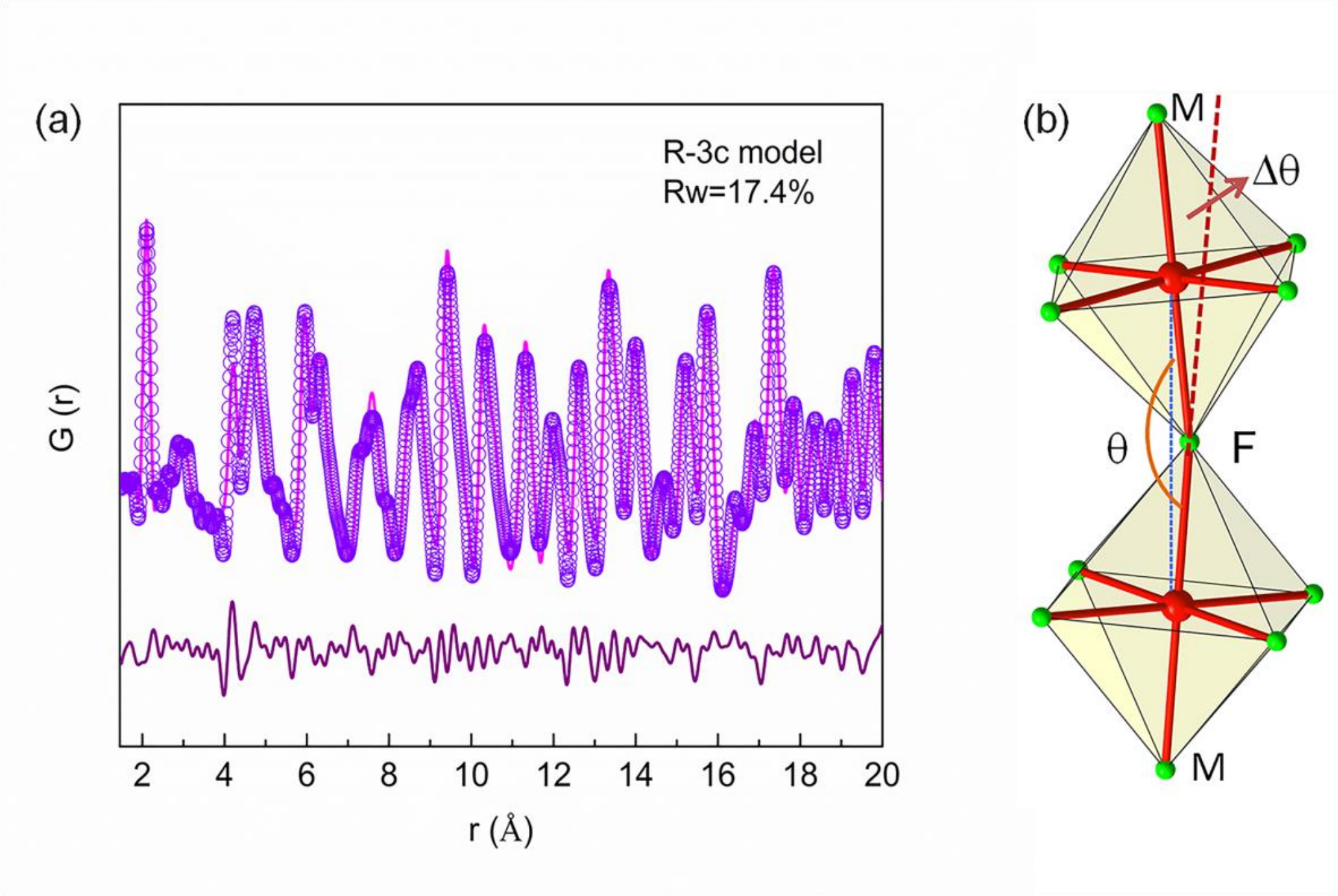

**Figure 3.** (a) Pair distribution function (PDF) fit of synchrotron X-ray scattering obtained at 300 K for $(Sc_{0.5}Ti_{0.5})F_3$ with the rhombohedral model at low r (1.7-20Å). The violet circles and red line correspond to the experimental and calculated data, respectively. Difference curve is shown by the purple line at the bottom. (b) The angle of *M*-F-*M* linkage, θ, extracted from the rhombohedral model which was used to fit to the PDF data of $(Sc_{1-x}M_x)F_3$ system ranging from1.7–20 Å.

In order to elucidate the correlation between the behavior of thermal expansion and local lattice distortion, Fig. 4 depicts the CTE data as a function of distortion degree of Δθ for $ScF_3$-based compounds and those representative trifluorides. The specific data are summarized in Table S1. Intriguingly, a close correlation is demonstrated clearly between local structure distortion and thermal expansion. With increasing local lattice distortion, NTE of $(Sc_{1-x}M_x)F_3$ becomes more and more reduced. In details, with the chemical substitution of $TiF_3$ for $ScF_3$, the CTE changes from $-3.4 \times 10^{-6}$ $K^{-1}$ for $ScF_3$ to $-1.51 \times 10^{-6}$ $K^{-1}$ for $(Sc_{0.1}Ti_{0.9})F_3$, of which the distortion degree simultaneously changes from 4.21° to 6.33°. To summarize, they are all well in accordance with this relationship including the present solid solutions of $(Sc_{1-x}Ti_x)F_3$, $(Sc_{0.9}Al_{0.1})F_3$, $(Sc_{0.9}Ga_{.1})F_3$, and the other previously reported compounds[17].

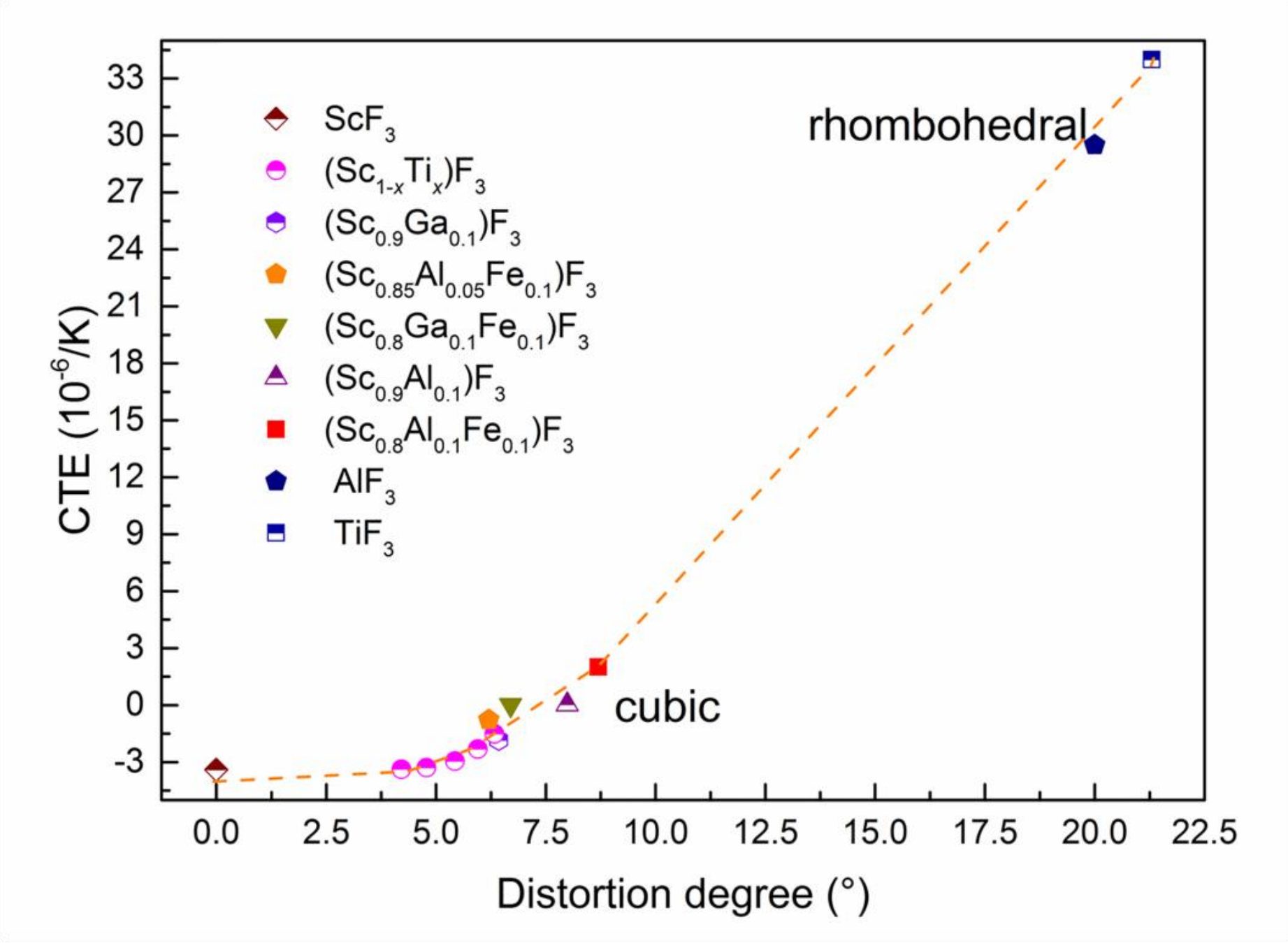

**Figure 4.** The correlation between local lattice distortion and CTE for $ScF_3$-based compounds and other representative trifluorides. The data of $(Sc_{0.85}Al_{0.05}Fe_{0.1})F_3$, $(Sc_{0.8}Ga_{0.1}Fe_{0.1})F_3$, $(Sc_{0.85}Al_{0.05}Fe_{0.1})F_3$ and $AlF_3$ are adopted from the literature.[17]

It can be expected that after the introduction of local lattice distortion transverse vibration of F atoms would contribute less to NTE but more to positive thermal expansion (PTE) component. The previous work has already claimed that the NTE behavior is favored by the straight M-X-M linkage (here X and M indicate anion and cation, respectively)[20]. And the bent M-X-M linkage tends to give rise to PTE. This is confirmed by rhombohedral metal fluorides showing very strong PTE (Fig. 4), in which the M-F-M linkage is much bent, like $TiF_3$ and $AlF_3$ (the angle of Ti-F-Ti is 157.9° and that of Al-F-Al is 159°)[17]. In the previous work of $Sc_{1-x}M_xF_3$ ($M$ = Ga, Fe; Al, Fe)[11,17], the controllable thermal expansion was also found to be correlated with the distortion of M-F-M linkages. Thus, controllable thermal expansion can be achieved if local lattice distortion is appropriately adjusted by different substitutions.

## 3. Conclusion

The solid solutions of $(Sc_{1-x}Ti_x)F_3$ ($x$ = 0.1, 0.3, 0.5, 0.7, and 0.9), $(Sc_{0.9}Al_{0.1})F_3$, and $(Sc_{0.9}Ga_{0.1})F_3$ were synthesized to study the correlation between local lattice distortion and thermal expansion. By the chemical substitution of $TiF_3$ for $ScF_3$, the thermal expansion behavior of $(Sc_{1-x}Ti_x)F_3$ becomes less and less negative. Different from cubic $ScF_3$, there are local lattice distortions in solid solutions of $(Sc_{1-x}Ti_x)F_3$, $(Sc_{0.9}Al_{0.1})F_3$, and $(Sc_{0.9}Ga_{0.1})F_3$. The relationship between thermal expansion and local lattice distortion is elucidated clearly. The larger local lattice distortion is, the more reduced NTE becomes. The present study could provide an effective method to control thermal expansion for those NTE materials with open-framework structure.

**Acknowledgements**

This work was supported by the National Natural Science Foundation of China (grant nos. 21322102, 91422301, 21231001, and 21590793), the Changjiang Young Scholars Award, National Program for Support of Top-notch Young Professionals, and the Fundamental Research Funds for the Central Universities, China (FRF-TP-14-012C1). This research used resources of the Advanced Photon Source, a U.S. Department of Energy (DOE) Office of Science User Facility operated for the DOE Office of Science by Argonne National Laboratory under Contract No. DE-AC02-06CH11357.

# Supporting Information for "Negative thermal expansion and local lattice distortion in the $(Sc_{1-x}Ti_x)F_3$ and related solid solutions"

## Experimental section

The samples were prepared via a solid state synthesis route. The raw materials of $ScF_3$ (99.99%), $TiF_3$ (98%), $AlF_3$ (99.9%), $GaF_3$ (99.99%) and $NH_4F$ (99.99%) powders, were mixed according to the stoichiometric starting reagents of $(Sc_{1-x}Ti_x)F_3$, $(Sc_{0.9}Al_{0.1})F_3$, and $(Sc_{0.9}Ga_{0.1})F_3$. These mixtures were pressed into small pellets which were then sandwiched into additional $NH_4F$ powder. Then the samples were sealed carefully into Cu tubes in order to avoid oxidization. Subsequently, these samples were heated at 850 °C for 2 h under ambient condition, before they were gradually cooled to room temperature. High-energy synchrotron X-ray diffraction (SXRD) data were collected over an angular range of 1-8° at the beamline 11-ID-C at the Advanced Photon Source (APS). The X-ray wavelength was λ = 0.117418 Å. The synchrotron X-ray scattering data for pair distribution function (PDF) were also collected at the same beamline of 11-ID-C of APS with the identical X-ray wavelength (λ = 0.117418 Å). All these X-ray scattering data were analyzed using the software of PDFgetX3,[1] and structure refinements were achieved by PDFgui.[2] Temperature-dependence of XRD data were collected from 300 to 900 K using a laboratory X-ray diffractometer (PANalytical, PW 3040-X'PertPro). The cubic ($Pm\bar{3}m$) model was adopted for the Rietveld refinement based on *Fullprof* software.

## Results and discussion

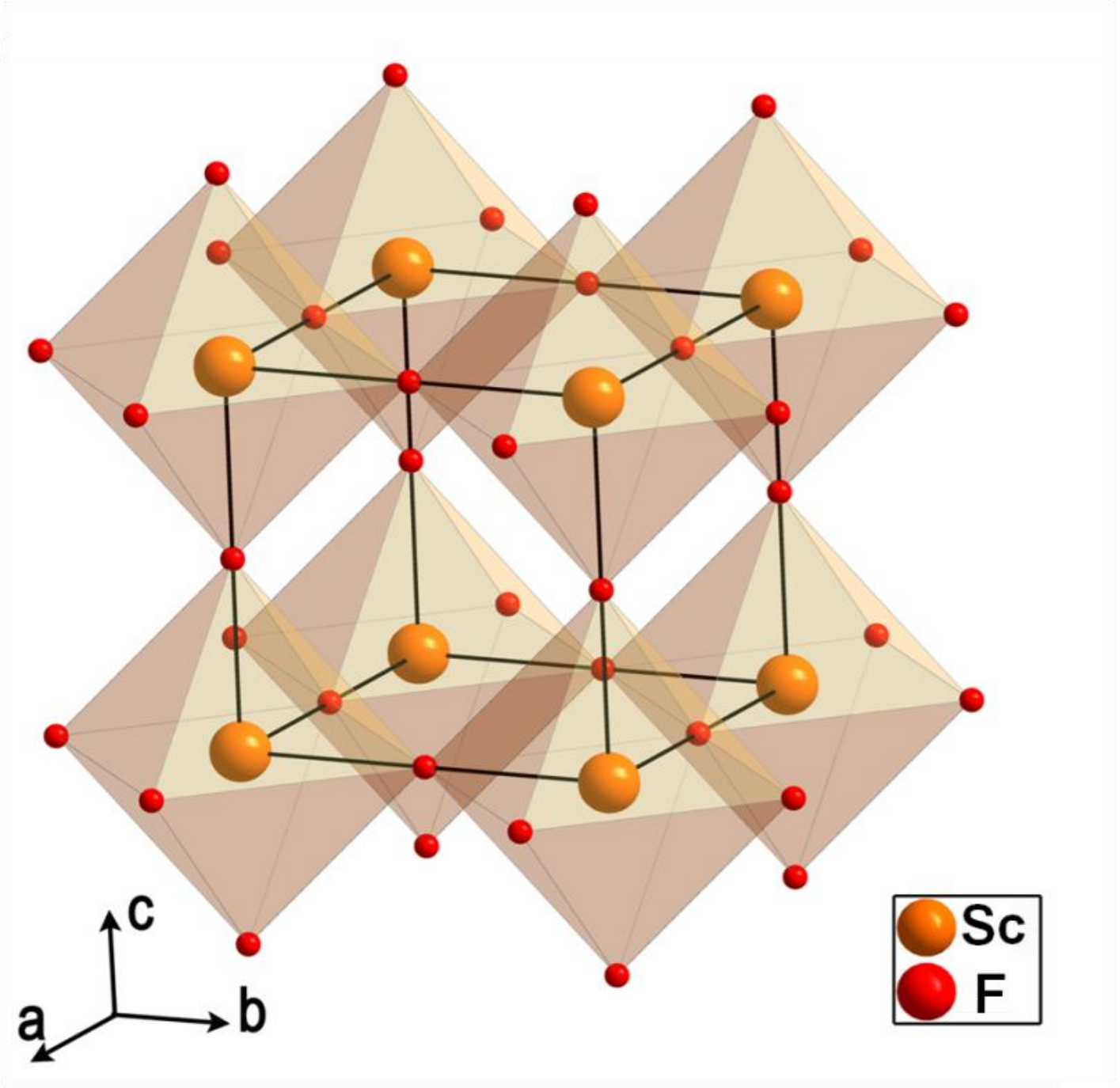

**Figure S1.** Crystal structure of cubic $ScF_3$.

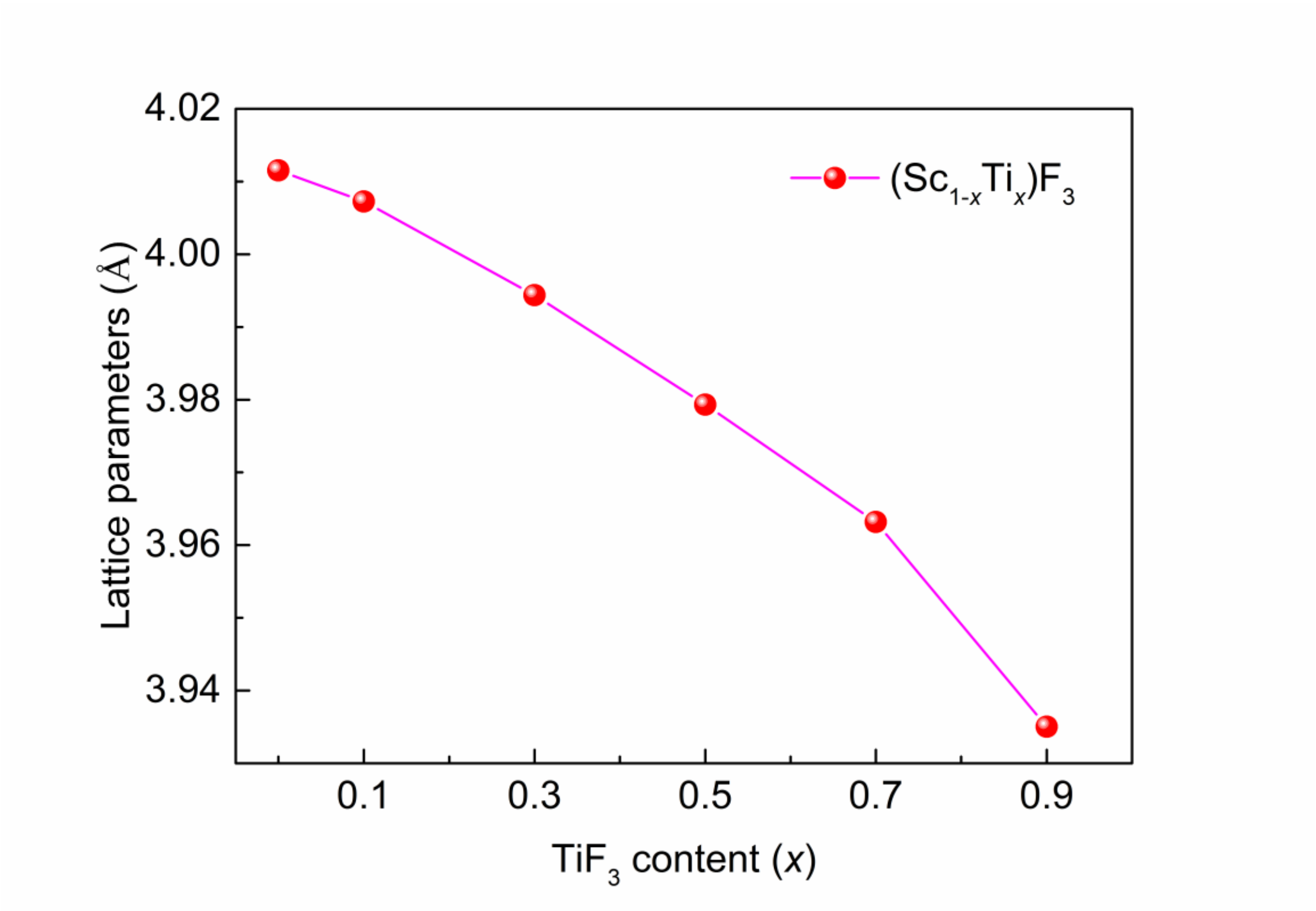

**Figure S2.** Lattice parameters of $(Sc_{1-x}Ti_x)F_3$ ($x$ = 0, 0.1, 0.3, 0.5, 0.7 and 0.9) solid solutions at room temperature.

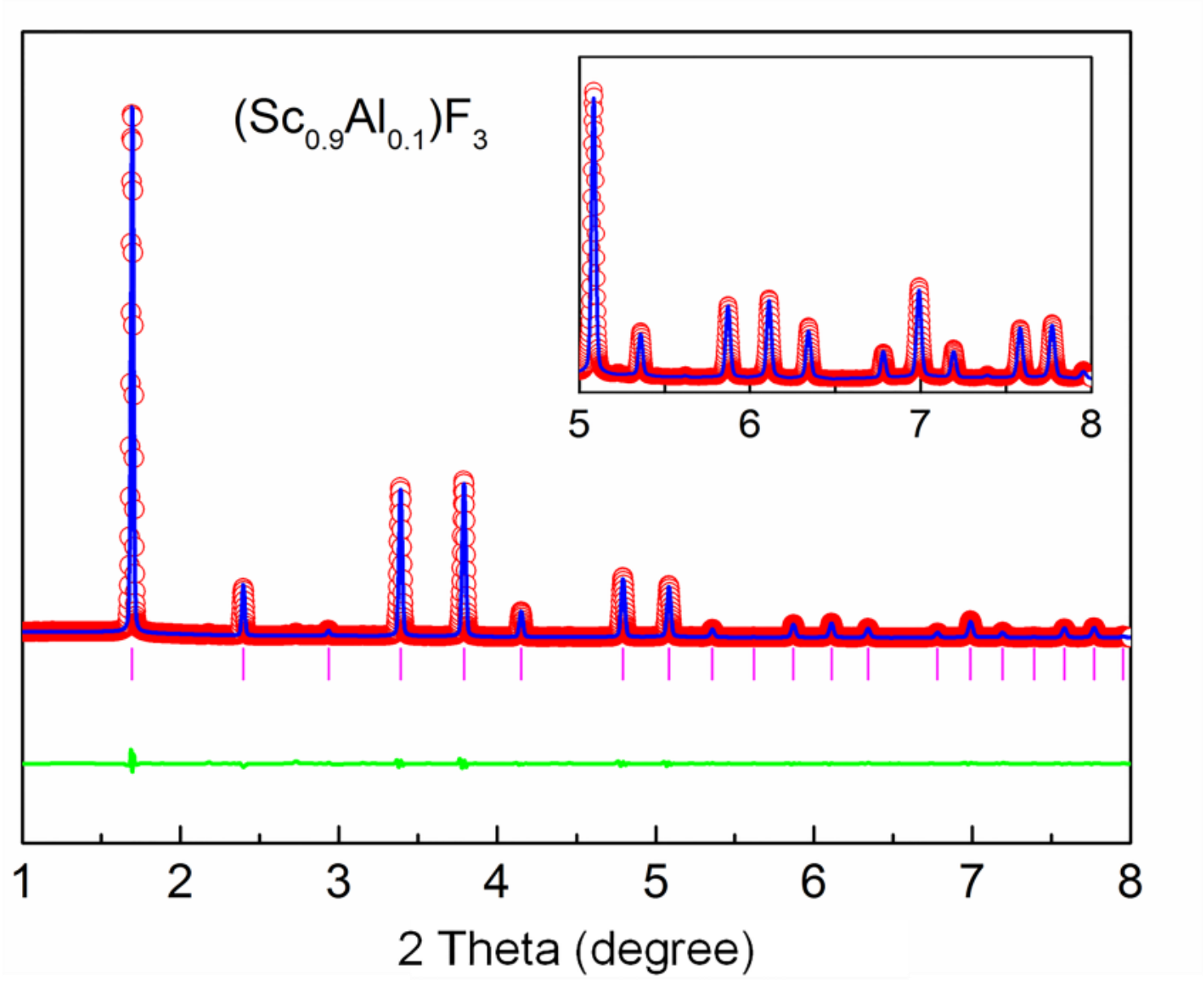

**Figure S3.** Observed, calculated, difference patterns of structure refinements of high-energy synchrotron XRD data of $(Sc_{0.9}Al_{0.1})F_3$.

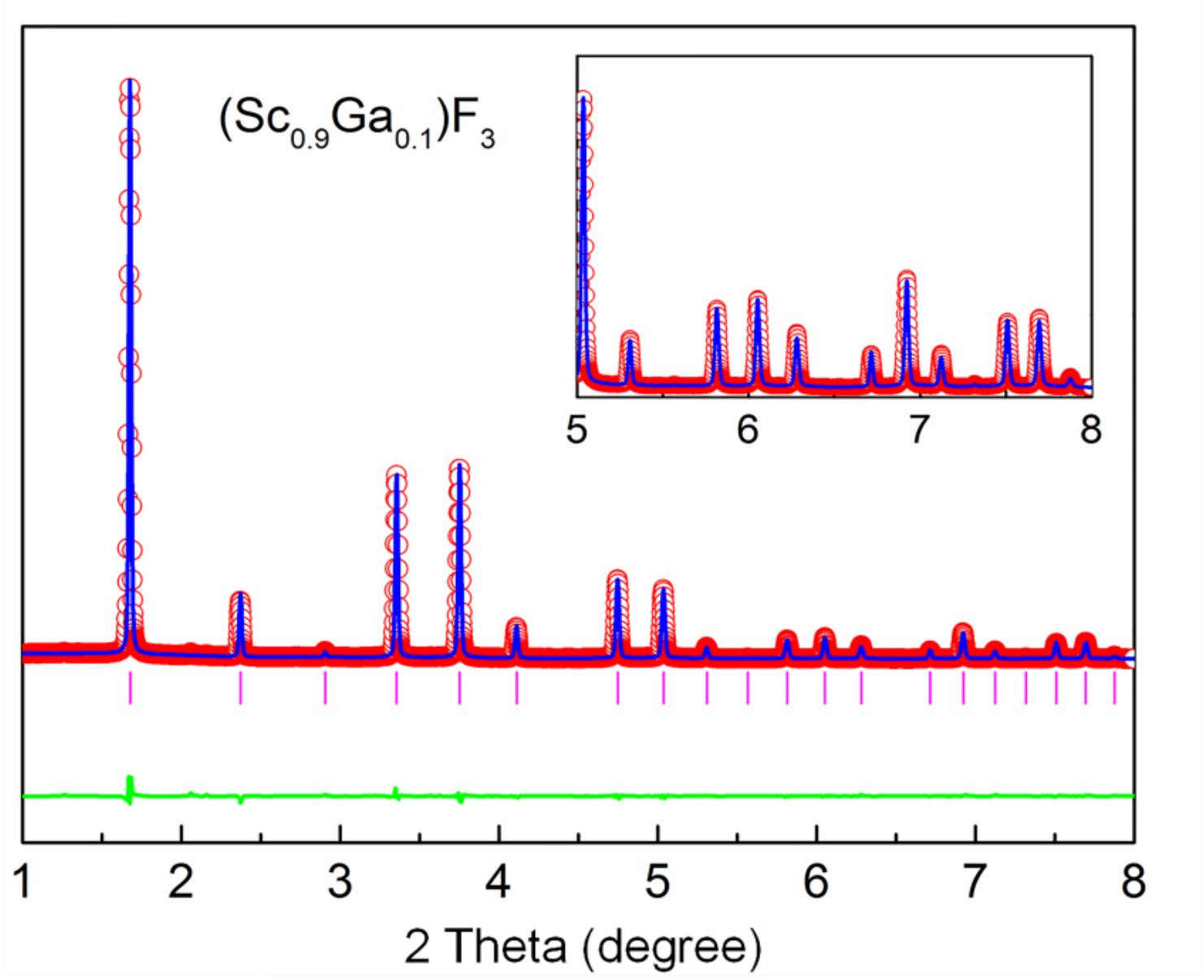

**Figure S4.** Observed, calculated, difference patterns of structure refinements of high-energy synchrotron XRD data of $(Sc_{0.9}Ga_{0.1})F_3$.

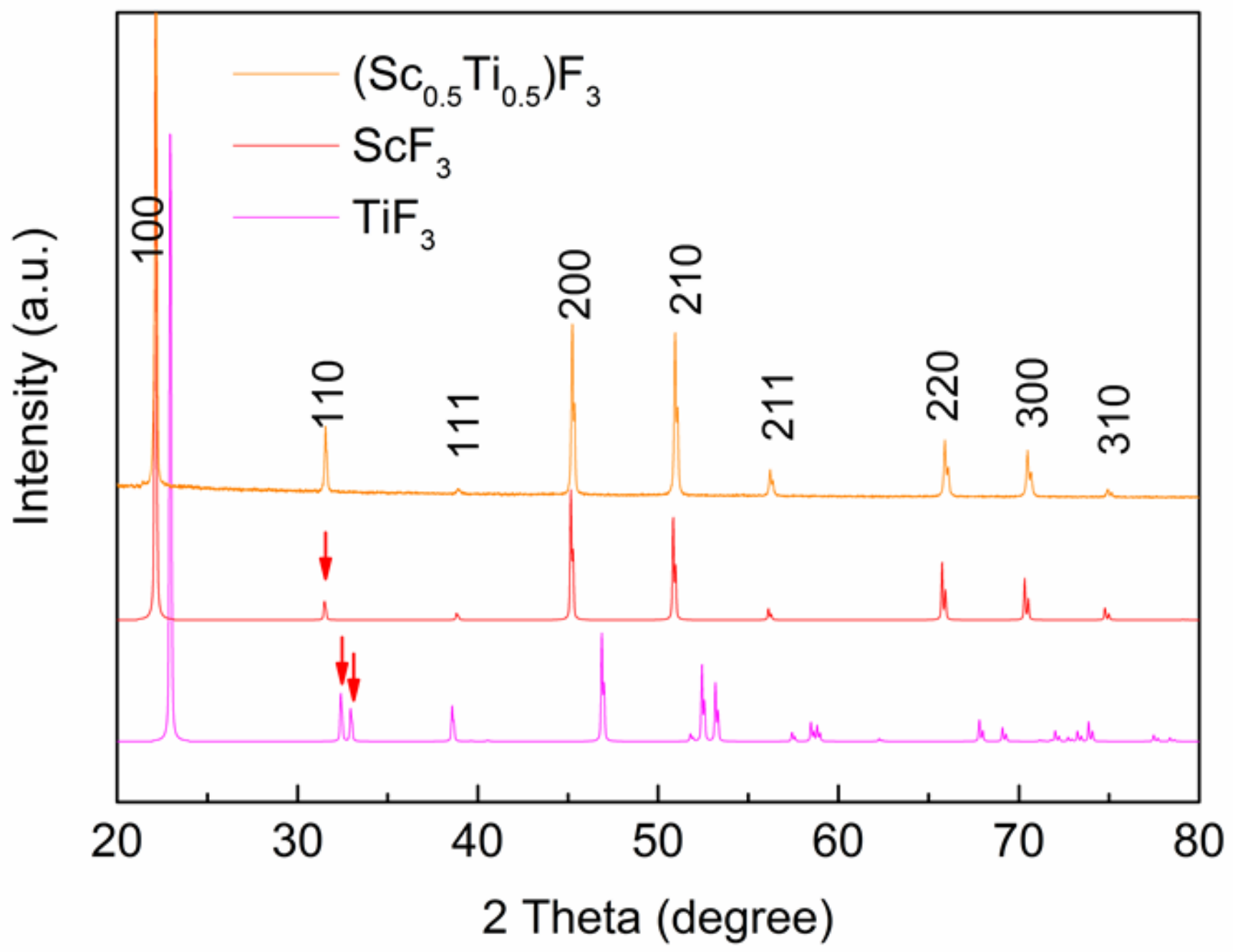

**Figure 5.** The XRD patterns of the $(Sc_{0.5}Ti_{0.5})F_3$ sample with $R\bar{3}c$ $TiF_3$ and $Pm\bar{3}m$ $ScF_3$ standard cards.

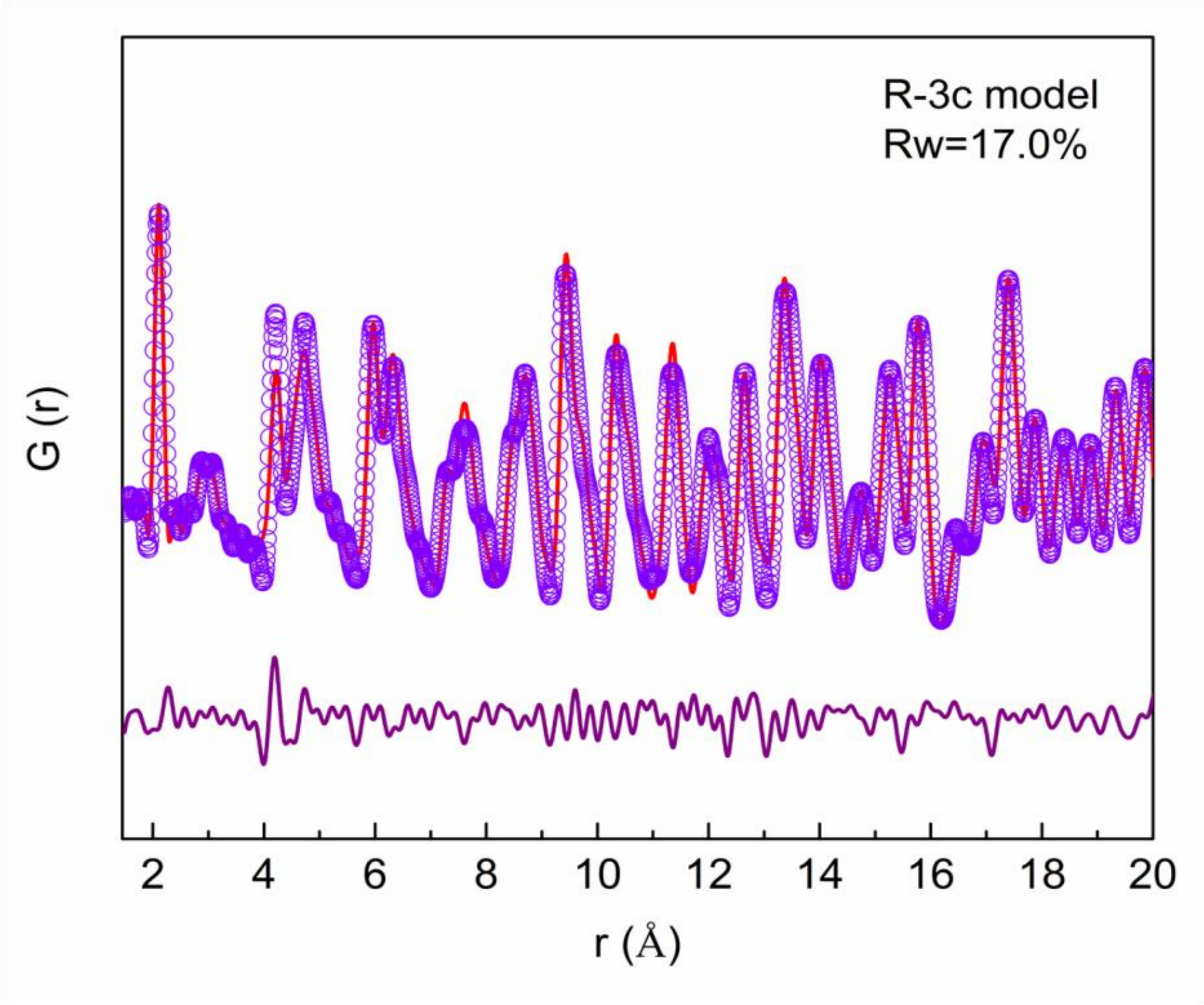

**Figure S6.** Pair distribution function (PDF) fit of synchrotron X-ray scattering obtained at room temperature for $(Sc_{0.9}Al_{0.1})F_3$ with the rhombohedra model at low r (1.7-20 Å). The violet circles and red line correspond to the experimental and calculated data, respectively. Difference curve is shown by the purple line at the bottom.

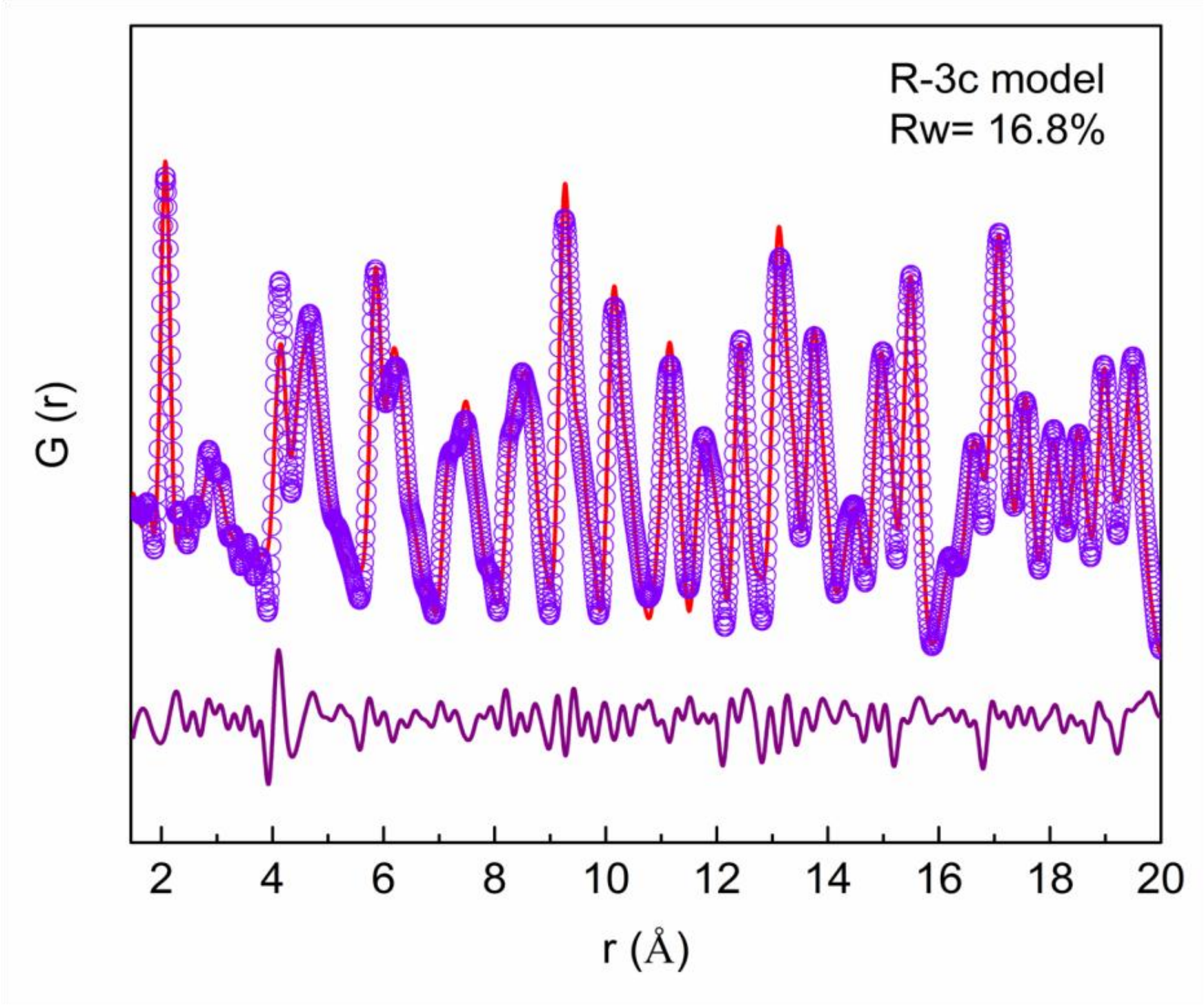

**Figure S7.** Pair distribution function (PDF) fit of synchrotron X-ray scattering obtained at room temperature for $(Sc_{0.9}Ga_{0.1})F_3$ with the rhombohedra model at low r (1.7-20 Å). The violet circles and red line correspond to the experimental and calculated data, respectively. Difference curve is shown by the purple line at the bottom.

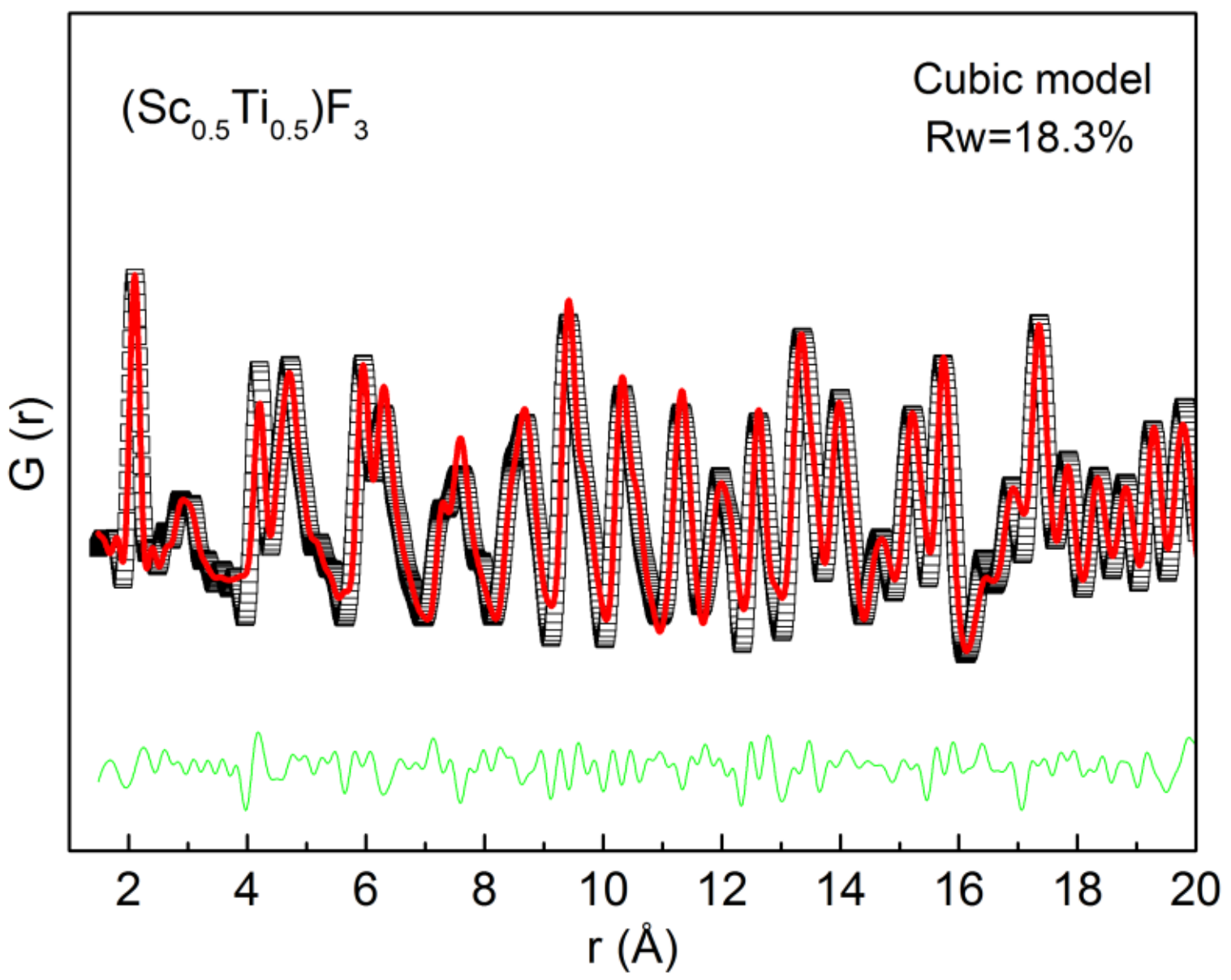

**Figure S8.** Pair distribution function (PDF) fit of synchrotron X-ray scattering G(r) functions for $(Sc_{0.5}Ti_{0.5})F_3$ with cubic model. The black squares and the red line represent the observed data and the fitted one, respectively. The green line at the bottom indicates the difference between the observed and fitted data.

**Table S1.** Thermal expansion and lattice distortion angles for $ScF_3$-based compounds and $TiF_3$.

| Compounds | Angle of M-F-M (θ) | Distortion degree (Δθ) | Linear CTE ($10^{-6}$ $K^{-1}$) | References |
|---|---|---|---|---|
| $ScF_3$ | 180 ° | 0 ° | -3.4 | Ref. 3 |
| $(Sc_{0.9}Ti_{0.1})F_3$ | 175.8 ° | 4.2 ° | -3.37 | this study |
| $(Sc_{0.7}Ti_{0.3})F_3$ | 175.2 ° | 4.8 ° | -3.29 | this study |
| $(Sc_{0.5}Ti_{0.5})F_3$ | 174.6 ° | 5.4 ° | -2.94 | this study |
| $(Sc_{0.3}Ti_{0.7})F_3$ | 174.1 ° | 6.0 ° | -2.31 | this study |
| $(Sc_{0.1}Ti_{0.9})F_3$ | 173.7° | 6.3 ° | -1.51 | this study |
| $(Sc_{0.9}Ga_{0.1})F_3$ | 173.6 ° | 6.4 ° | -1.84 | this study |
| $(Sc_{0.9}Al_{0.1})F_3$ | 172.0 ° | 8.0 ° | -0.53 | this study |
| $TiF_3$ | 157.9 ° | 21.3 ° | 36.0 | Ref. 4 |

The temperature range of all $ScF_3$ based compounds is from 300 to 800 K. That of $TiF_3$ ranges from 10 to 375 K.